\documentclass[11pt]{article}
\usepackage[dvips]{color}                 
\usepackage{graphicx}                     
\usepackage{amssymb}
\usepackage{dsfont}
\usepackage{amsmath}
\renewcommand{\today}{\ifcase\day\or 1st\or 2nd\or 3rd\or 4th\or 5th\or 6th\or
 7th\or 8th\or 9th\or 10th\or 11th\or 12th\or 13th\or 14th\or 15th\or 
 16th\or 17th\or 18th\or 19th\or 20th\or 21st\or 22nd\or 23rd\or 24th\or
 25th\or 26th\or 27th\or 28th\or 29th\or 30th\or 
 31st\fi~\ifcase\month\or January\or February\or March\or April\or
 May\or June\or July\or August\or September\or October\or November\or
 December\fi \space \number\year}   
\newcommand{\mytitle}[1]{
                         \begin{center}
                           \LARGE{\textbf{#1}}
                         \end{center}}
\newcommand{\myauthor}[1]{\textbf{#1}}
\newcommand{\myaddress}[1]{\textit{#1}}
\newcommand{\mypreprint}[1]{\begin{flushright} #1 \end{flushright}}
\begin{document}

\begin{titlepage}
\mypreprint{
TUM-T39-06-13 \\
}

\vspace*{0.5cm}
\mytitle{The electromagnetic Nucleon to Delta transition in Chiral Effective
Field Theory}
  \vspace*{0.3cm}

\begin{center}
 \myauthor{Tobias A. Gail} and
  \myauthor{Thomas R. Hemmert}, 

  \vspace*{0.5cm}
\myaddress{Theoretische Physik T39, Physik Department\\
	TU-M\"unchen, D-85747 Garching, Germany}
  \vspace*{0.2cm}
\end{center}

\vspace*{0.5cm}

\begin{abstract}
\noindent
We present a calculation of 
the three complex form factors parametrizing the nucleon to $\Delta$
transition matrix element in the framework of chiral effective field
theory with explicit $\Delta$ degrees of freedom. The interplay
between short and long range physics is discussed and estimates for
systematic uncertainties due to higher order effect are given.
\end{abstract}
\end{titlepage}

\vfill\pagebreak

\section{Introduction}

In this talk we discuss the findings of reference
\cite{GH}, where the low energy behaviour of the nucleon to $\Delta$
electromagnetic transition ($N(p_N)+\gamma^*(q)\rightarrow \Delta(p_{\Delta})$) was analyzed in the framework of chiral effective field
theory. We give an overview of the theoretical tools which are utilized
in such an analysis in the next section and discuss the results in the
third paragraph.\\
Demanding Lorentz covariance, gauge invariance and parity conservation
the matrix element of a vector $I\left(J^P\right)=\frac{3}{2}\left(\frac{3}{2}^+\right)$ to 
$\frac{1}{2}\left(\frac{1}{2}^+\right)$ transition (like
$N\gamma^*\rightarrow \Delta$) can be parametrized
in terms of three form factors, i.e. complex valued functions of the
momentum transfer squared. For our calculation we follow the conventions of ref.\cite{GHKP} and choose the definition:
\begin{eqnarray}
i\mathcal{M}_{\Delta\rightarrow N\gamma} & = &\sqrt{\frac{2}{3}}
\,\frac{e}{2M_N}\bar{u}(p_N)\gamma_5\bigg[G_1(q^2)
(\not\!q\epsilon_{\mu}-\not\!\epsilon
q_{\mu})+\frac{G_2(q^2)}{2M_N}(p_N\cdot\epsilon
q_{\mu}-p_N\cdot q\epsilon_{\mu}) \nonumber
\\&&+\frac{G_3(q^2)}{2\Delta}(q\cdot
\epsilon q_{\mu}-q^2\epsilon_{\mu})\bigg]u^{\mu}_{\Delta}(p_{\Delta}).
\label{defff}
\end{eqnarray}
Here $e$ denotes the charge of the electron, $M_N$ is the mass of a
nucleon and $\Delta=M_{\Delta}-M_N$ the $\Delta$-nucleon mass splitting, $p_{N/\Delta}^{\mu}$ denotes the
relativistic four-momentum of the outgoing nucleon/incoming $\Delta$ and $(q^{\mu}$,
$\epsilon^{\mu})$ are the momentum and polarization vectors of the outgoing photon,
respectively. From the point of view of chiral effective field theory the
signatures of chiral dynamics in the $N\Delta$-transition are particularly transparent in the $G_i(q^2),\,i=1,2,3$ basis, which serves as the 
analogue of the
Dirac- and Pauli-form factor basis in the  vector current of a nucleon. However, most experiments and most model calculations refer to the
multipole basis of the general $N\Delta$-transition current, i.e.
the magnetic dipole $\mathcal{G}_M^*(Q^2)$, electric quadrupole
$\mathcal{G}_E^*(Q^2)$ and Coulomb quadrupole $\mathcal{G}_C^*(Q^2)$
form factor defined by Jones and Scadron \cite{JS}\footnote{The relation
between both sets can be found in \cite{GH}.}. Phenomenological
information about  the electromagnetic nucleon to $\Delta$ transition is gained in the process $e\,p\rightarrow e^\prime\,N\pi$ in the region of the $\Delta$-resonance 
({\it e.g.} see ref.\cite{OOPS} and references given therein), which has access to a lot more hadron structure properties than just the 
$N\Delta$-transition current of eq.(\ref{defff}). However, we compare
our results to experimental data assuming the approximate relations
\begin{eqnarray}
\textnormal{EMR}(q^2) & \equiv &Re\left[\frac{E_{1+}^{I=3/2}(W_{res},q^2)}{M_{1+}^{I=3/2}(W_{res},q^2)}\right] \approx
	-\textnormal{Re}\left[\frac{\mathcal{G}_E^*(q^2)}{\mathcal{G}_M^*(q^2)}\right], \label{EMR} \\
\textnormal{CMR}(q^2) & \equiv
&Re\left[\frac{S_{1+}^{I=3/2}(W_{res},q^2)}{M_{1+}^{I=3/2}(W_{res},q^2)}\right]\approx
	\nonumber \\ &  &
	-\frac{\sqrt{((M_{\Delta}+M_N)^2-q^2)((M_{\Delta}-M_N)^2-q^2)}}{4M_{\Delta}^2}
	\textnormal{Re}\left[\frac{\mathcal{G}_C^*(q^2)}{\mathcal{G}_M^*(q^2)}\right].\label{CMR}
\end{eqnarray} 
Concerning the resonance pole contributions alone, the right hand side of the above equations represents the
 ratios at the T-matrix pole $M_{\Delta}=(1210-i\,50)$MeV
\cite{PDG}. In this work we assume that the form factor ratios are approximately equal to the outcome of the
various data analyses of the electroproduction multipoles
$M_{1+}^{I=\frac{3}{2}}$, $E_{1+}^{I=\frac{3}{2}}$ and
$S_{1+}^{I=\frac{3}{2}}$. Ultimately the validity of this (approximate) connection between the pion-electroproduction multipoles and the 
$N\Delta$-transition form factors has to be
checked in a full theoretical calculation.

\section{Theoretical Framework}

\begin{figure}
\begin{center}
\includegraphics[width=12cm,clip]{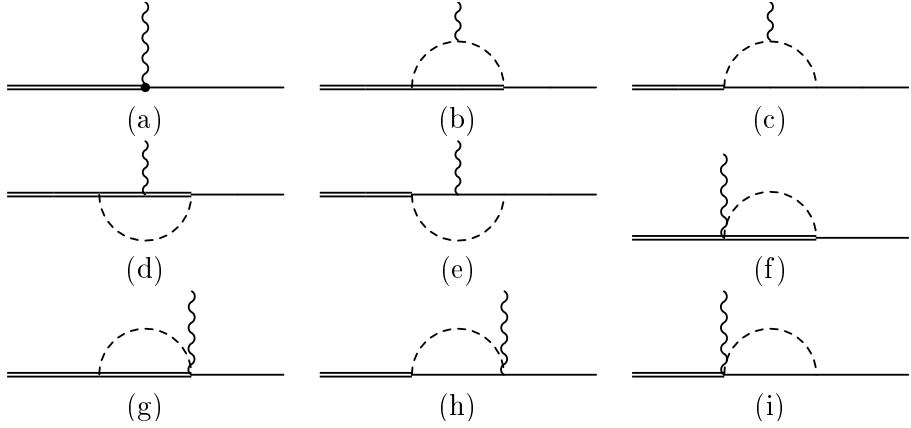}
\caption{The diagrams contributing to the $\Delta\rightarrow N\gamma$ transition at leading-one-loop
order in the SSE formalism \cite{GHKP}. The single solid lines
represent nucleon-,the double lines $\Delta$-propagation. The dashed
lines symbolize intermediate pion fields and the wiggly lines stand
for the outgoing photon.}
\label{fig:feyn}
\end{center}
\end{figure}

In this section we briefly introduce the theoretical tools necessary to
calculate the $N\Delta$-matrix element eq.(\ref{defff}) at low
energies. In reactions with small momentum transfer (typically
$Q^2=-q^2\leq0.2$GeV$^2$) the nucleon can - due to a separation of
scales (the pion is much lighter than the next heavier hadron) - clearly be divided into a long-ranging pion cloud and a
small nucleon core. Chiral Effective Field Theory in the baryon sector
(ChPT, for a classic
paper see \cite{BKKM}) provides a
systematic framework for the calculation of pion cloud dynamics and
at the same time also encodes short range physics via local operators of (theoretically)
undetermined strength. A formulation of ChPT suitable for the
calculation of the $N\Delta$-transition is the ``small scale
expansion'' (SSE) \cite{HHK}, which includes explicit $\Delta$ degrees of
freedom and provides an expansion scheme for the chiral Lagrangean. This framework contains three
light (the momentum transfer $|Q|$, the pion mass $m_{\pi}$
and the $\Delta$-nucleon mass splitting $\Delta$) and two heavy (the
nucleon mass $M_N$ and the scale of chiral symmetry breaking $4\pi
F_{\pi}$, where $F_{\pi}$ is the pion decay constant) scales and each
ratio of a light to a heavy scale is counted as a small parameter of
the same order: $\epsilon\in\{\frac{|Q|}{4\pi F_{\pi}},\frac{m_{\pi}}{4\pi F_{\pi}},\frac{\Delta}{4\pi F_{\pi}},\frac{|Q|}{M_N},\frac{m_{\pi}}{M_N},\frac{\Delta}{M_N}\}$.\\
The Lagrangean containing all terms necessary for a leading one loop
order calculation (i.e. order $\epsilon^3$) reads \cite{GHKP,HHK}:
\begin{eqnarray}
\mathcal{L}_{SSE} & = & \mathcal{L}_{\pi\pi}^{(2)}+ \mathcal{L}_{\pi N}^{(1)}
+\mathcal{L}_{\pi\Delta}^{(1)}+\mathcal{L}_{\pi
N\Delta}^{(1)}+\mathcal{L}_{\gamma N\Delta}^{(2)}+\mathcal{L}_{\gamma
N\Delta}^{(3)}.
\label{Lag}
\end{eqnarray}
At leading order $\mathcal{L}_{\pi\pi}^{(2)}$ encodes the pion
dynamics, while $\mathcal{L}_{\pi N}^{(1)}$ and
$\mathcal{L}_{\pi\Delta}^{(1)}$ describe the pion-nucleon and the
pion-$\Delta$ system respectively and $\mathcal{L}_{\pi
N\Delta}^{(1)}$ contains the pion-nucleon-$\Delta$ coupling. These
parts of the Lagrangean form the input (i.e. vertizes and
propagators) for the leading one loop calculation. All
Feynman diagrams of a photon field coupling to the $\pi N \Delta$
system contributing at order $\epsilon^3$ in the SSE scheme are shown
in figure \ref{fig:feyn}.\\
To the order we are working, the Lagrangeans $\mathcal{L}_{\gamma N\Delta}^{(2)}$ and
$\mathcal{L}_{\gamma N\Delta}^{(3)}$ contribute only to the values of
$G_1(0)$ and $G_2(0)$ at zero momentum transfer through local
operators (represented by diagram (a) in figure \ref{fig:feyn}). The strength of these operators is
undetermined in the effective field theory approach and will be extracted from phenomenology in the next
section. Our analysis \cite{GH} has shown\footnote{The reasoning given
there is based on the following two main arguments: \textbf{1.} The
$Q^2$-dependence  of $G_2$ resulting from the pion cloud alone is considered
to be unphysical. We expect a form factor to tend to zero for large
momentum transfer, which $G_2(Q^2)$ can only fulfill if a radius
correction of adequate size is included. \textbf{2.} The impact of this radius
correction is disproportionally magnified by the translation into the
Jones-Scadron form factor basis and therefore has already to be
included in an analysis of these form factors  at order $\epsilon^3$. A numerical discussion of this effect
is given in the next section.}, that the local
operator which gives the contributions from the nucleon core to the form
factor $G_2(Q^2)$ a finite spacial extension (i.e. a contribution to
the radius of this form factor which formally is of higher order) is underestimated by the strict counting rules and its
inclusion extends the $Q^2$-range of applicability of the leading
one loop calculation significantly. The dependence of the quantities $\mathcal{G}_M^*(Q^2)$, EMR$(Q^2)$ and CMR$(Q^2)$
on the form factors $G_1(Q^2)$, $G_2(Q^2)$ and $G_3(Q^2)$ is such, that this radius contribution only has a very small impact in the magnetic dipole form
factor (which is highly dominated by $G_1(Q^2)$), but has decisive
influence in the quadrupole moments. Altogether we consider short range
contributions to the static limit ($q^2=0$) of $G_1$ and $G_2$ and the radius of
$G_2$ in our calculation which thus contains three free parameters. All
coupling constants contained in the other parts of the Lagrangean are
extracted from other observables\footnote{For the full expressions for the
Lagrangean eq.(\ref{Lag}) and the values of the coupling constants
appearing therein see reference \cite{GH}.}.\\
Besides the analysis presented here the authors of reference \cite{PV}
also performed an effective field theory calculation of the
$N\Delta$-transition at leading one loop level. Let us briefly point out the differences between both ChPT
calculations: 
\begin{enumerate}
\item In reference \cite{PV} a different expansion scheme, namely the
$\delta$-scheme \cite{d-scheme} has been applied to introduce a hierarchy of terms. In this
scheme not all ratios of a small to a large scale are counted as a small
parameter of the same order -- as it is done in the $\epsilon$-counting
discussed above -- but the small
parameters are: $\delta^2 \in
\{\frac{\Delta^2}{(4\pi F_{\pi})^2},\frac{m_{\pi}}{4\pi
F_{\pi}},\frac{|Q|^n}{(4\pi F_{\pi})^n}\}$. Thus the pion mass $m_{\pi}$
is counted to be of the order of the $\Delta$-nucleon mass-splitting
squared. The order $n$ of the momentum transfer then depends on its
typical size in the particular process: if $|Q|\approx m_{\pi} \Rightarrow n=2$ and if
$|Q|\approx \Delta  \Rightarrow n=1$. The different counting
leads to a different estimate of the relevance (i.e. inclusion or
omission at a certain order) of the Feynman diagrams contributing to
the $N$ to $\Delta$ matrix element between the SSE and the $\delta$-expansion. Diagram (b) of figure
\ref{fig:feyn}, for instance, provides important structures for our
result, while it is neglected in the $\delta$-counting approach at leading one
loop level.    
\item While we performed a non-relativistic expansion of the Lagrangean
(i.e. not only considered the chiral symmetry breaking scale $4\pi
F_{\pi}$ but also the nucleon mass $M_N$ as a large scale) reference
\cite{PV} gives a covariant calculation (i.e. a resummation of all
terms which are suppressed by any power of the nucleon mass at a
certain order of the chiral symmetry breaking scale). The results
show that the inclusion of all terms which are suppressed by
inverse powers of the nucleon mass cancels out important parts of the
curvature in the  $Q^2$-dependence of the resulting form factors. The
same effect has already been observed in the vector current of the
nucleon \cite{BFHM,KM}. On
the other hand the additional string of $\frac{m_{\pi}}{M_N}$-terms provides enough
quark mass dependent structures to qualitatively describe their quark
mass dependence up to currently available lattice data \cite{Alexandrou}.
\item Each of the two studies includes short range physics in a
different manner: In \cite{GH} we took into account the accordant contributions to
$G_1(0)$ and $G_2(0)$ arising from the chiral Lagrangean and furthermore found short range physics to play an
important role in the radius of the $G_2$ form factor. The authors of reference
\cite{PV} deal with three {\em structurally different} free parameters: the coupling constants
characterizing the magnetic dipole and the electric- and Coulomb
quadrupole transitions. Furthermore, they effectively include vector-meson exchange by giving the magnetic
dipole coupling a dipole-shape $Q^2$-dependence.\\
Finally the extraction of the free parameters from different experimental data
leads to further differences between the numerical results of both
calculations.

\end{enumerate}

\section{Discussion of the Results}

\begin{figure}
\begin{center}
\includegraphics[width=12cm,clip]{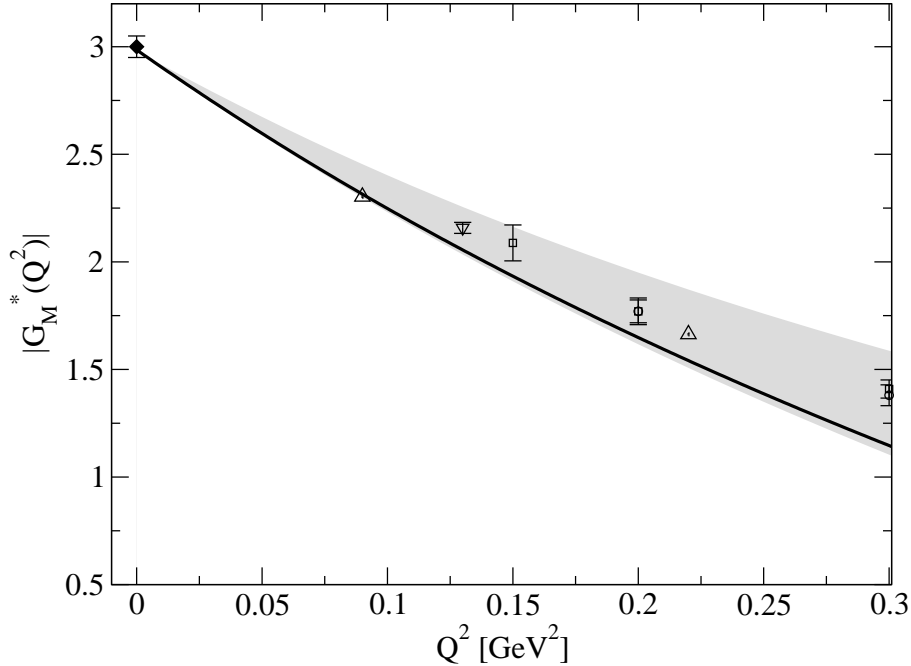}
\caption{The momentum transfer dependence of the magnetic dipole form
factor $\mathcal{G}_M^*(Q^2)$: The solid line represents the
$\mathcal{O}(\epsilon^3)$ SSE result \cite{GH} as discussed in the text, the
gray shaded band indicates the uncertainty of this result when neglecting
higher order effects. Data points are taken  from \cite{Beck} (diamond),
 \cite{Stein} (triangle up), \cite{rean} (triangle down), \cite{Baetzner} (square) and \cite{Bartel} (circle).}
\label{GM_err}
\end{center}
\end{figure}

\begin{figure}
\begin{center}
\includegraphics[width=12cm,clip]{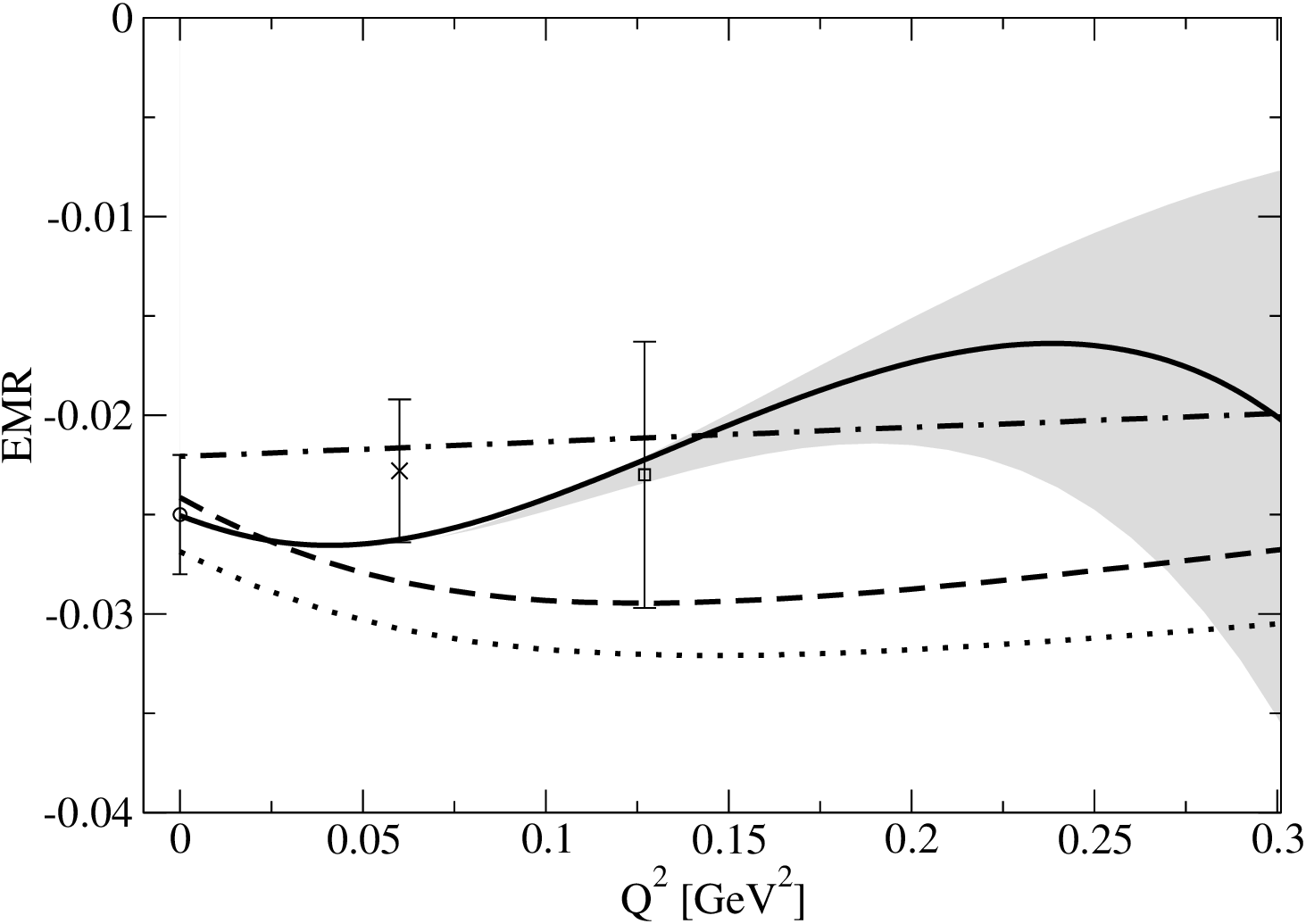}
\caption{Momentum transfer dependence of the ratio of the electric
quadrupole to the magnetic dipole form factor EMR$(Q^2)$: The solid line represents the
$\mathcal{O}(\epsilon^3)$ SSE result \cite{GH} as discussed in the text, the
gray shaded band indicates the uncertainty of this result when neglecting
higher order effects. The dashed-dotted (MAID \cite{DMTMAID}), dashed
(DMT \cite{DMTMAID}) and dotted (Sato-Lee \cite{SL}) curves are model
predictions. Experimental date are from MAMI (real photon
point \cite{Beck} and $Q^2=0.06$GeV$^2$ \cite{Stave})
and  OOPS \cite{OOPS} ($Q^2=0.127$GeV$^2$). }
\label{EMR_err}
\end{center}
\end{figure}

\begin{figure}
\begin{center}
\includegraphics[width=12cm,clip]{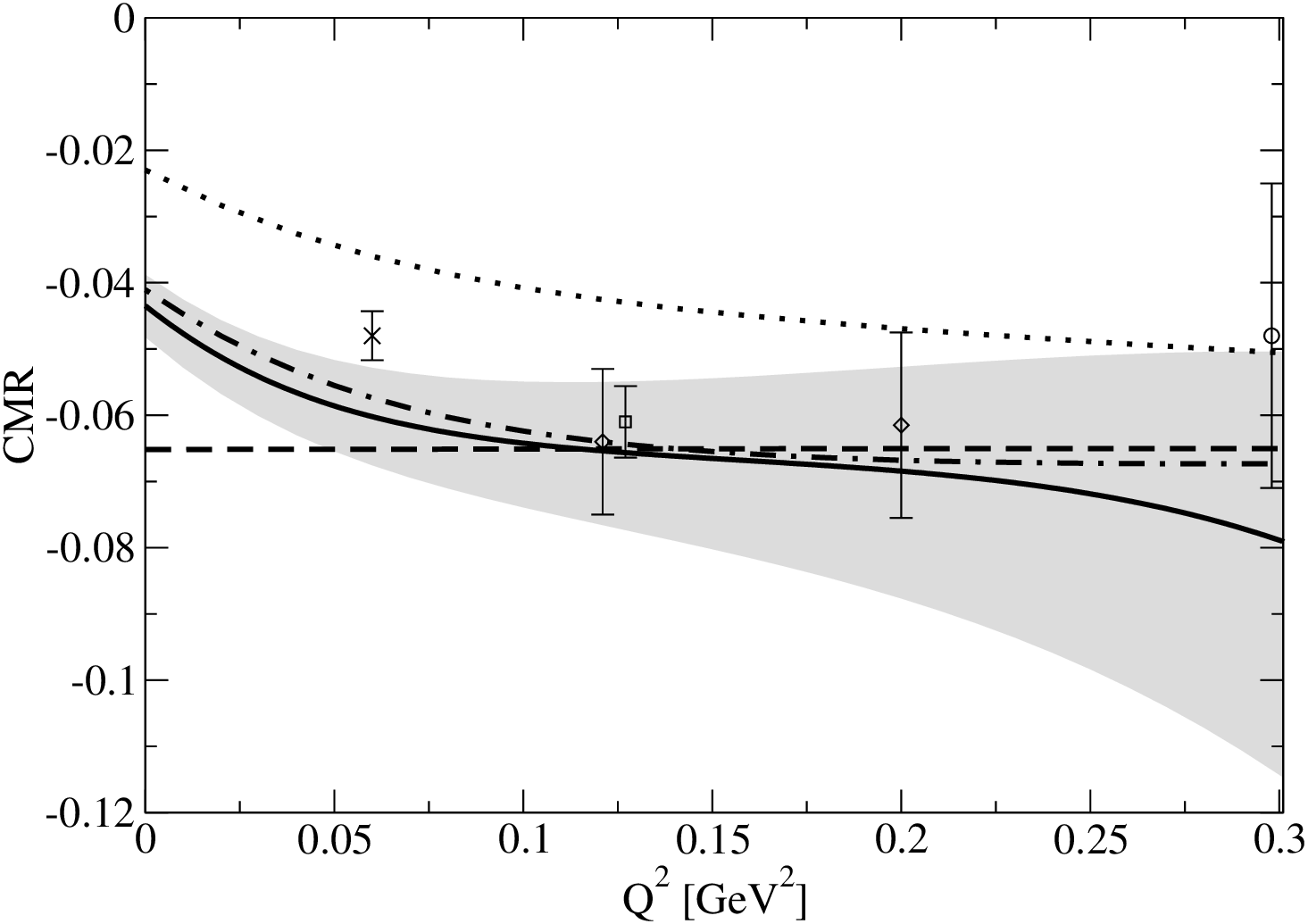}
\caption{Momentum transfer dependence of the ratio of the Coulomb
quadrupole to the magnetic dipole form factor CMR$(Q^2)$: The solid line represents the
$\mathcal{O}(\epsilon^3)$ SSE result \cite{GH} as discussed in the text, the
gray shaded band indicates the uncertainty of this result when neglecting
higher order effects. The dashed-dotted (MAID \cite{DMTMAID}), dashed
(DMT \cite{DMTMAID}) and dotted (Sato-Lee \cite{SL}) curves are model
predictions. The data-points shown are from refs.\cite{Stave} (cross), \cite{MAMI} (diamonds), \cite{Bonn} (circles) and \cite{OOPS} (square).}
\label{CMR_err}
\end{center}
\end{figure}

The analytic expressions of the transition
form factors resulting from a calculation in the SSE framework up to
order $\epsilon^3$ can be found in reference
\cite{GH}. In this section we give a brief overview of the findings
presented there and discuss
the size of the errors which arise as a consequence when neglecting
higher order effects. \\
The three free parameters contained in our calculation are extracted
from experimental data for the magnetic dipole form factor
$\mathcal{G}_M^*(Q^2)$ at $Q^2<0.2$GeV$^2$ and the value of EMR at the
real photon point $Q^2=0$.
Having fixed the low energy constants we arrive at the numerical
values given in  Table \ref{tab:radii}, where the real and
imaginary parts of the static limit values and the radii defined through
\begin{eqnarray}
G_i(Q^2) & = &
\textnormal{Re}[G_i(0)]\left[1-\frac{1}{6}r_{i,\textnormal{Re}}^2Q^2+...\right]+
i\,\textnormal{Im}[G_i(0)]\left[1-\frac{1}{6}r_{i,\textnormal{Im}}^2Q^2+...\right]
\label{defr}
\end{eqnarray}
are shown for both sets of form factors discussed in this work.
It is worth mentioning that the real parts of the radii of both the electric and the Coulomb $N\Delta$-transition form factors are negative!\\
\begin{table}
\begin{center}
\begin{tabular}{|c|cc|cc|cc|}
\hline
  & $\textnormal{Re}[G_i(0)]$ & $r_{i,\textnormal{Re}}^2$
[fm$^{2}$] & $\textnormal{Im}[G_i(0)]$
 & $r_{i,\textnormal{Im}}^2$ [fm$^{2}$] &
 $\left|G_i(0)\right|$ & $r_{i,\textnormal{Abs}}^2$ [fm$^{2}$] \\
\hline
$G_1$ & 4.95& 0.679 &0.216 & 3.20 & 4.96 & 0.678 \\
$G_2$ & 5.85& 3.15 & -10.0 & 1.28 & 11.6 & 1.73 \\
$G_3$ & -2.28& 3.39 & 2.01 & -2.26& 3.04 & 0.907 \\
\hline
$\mathcal{G}_M^*$ & 2.98 & 0.627 &-0.377 & 1.36 & 3.00 & 0.630 \\
$\mathcal{G}_E^*$ & 0.0441&-0.836 &-0.249 & 0.422& 0.253 & 0.388\\
$\mathcal{G}_C^*$ & 1.10& -0.729 &-1.68 & 1.90 & 2.01 & 1.10 \\
\hline
\end{tabular}
\caption{The values of the form factors defined in eq.(\ref{defff})
and the Jones-Scadron form factors \cite{JS} at $Q^2=0$ and their
radii \cite{GH}.}
\label{tab:radii}
\end{center}
\end{table}
While all imaginary parts occurring in our analysis are solely generated by
pion cloud effects, the numerical values for the  real parts result from an interplay between
short and long range physics. Separating their contributions to the
static limit one finds\footnote{Such a statement is of course renormalization
scale dependent. However, the qualitative statements remain correct for all
typical renormalization-scales of the system (0.6GeV$<\lambda$<1.5GeV)}:  
\begin{eqnarray}
\textnormal{Re}[\mathcal{G}_M^*(0)] |_{\lambda=1\,\textnormal{GeV}}& = &
\left.-1.06\right|_{pc}+\left.4.04\right|_{sd},
\label{ps1}\\
\textnormal{Re}[\mathcal{G}_E^*(0)] |_{\lambda=1\,\textnormal{GeV}} & = & \left.0.155\right|_{pc}-\left.0.110\right|_{sd},\\
\textnormal{Re}[\mathcal{G}_C^*(0)] |_{\lambda=1\,\textnormal{GeV}} & = & \left.1.47\right|_{pc}-\left.0.365\right|_{sd},
\label{ps2}
\end{eqnarray}
where `` sd'' marks contributions from short distance physics while
``pc''  labels pion cloud effects. The physical interpretation of this
is that the pion cloud strongly shields the characteristic of the nucleon core. Qualitatively the same statement can
be made about the form factor basis defined in eq.(\ref{defff}).  An
analogous effect has
e.g. already been observed in the anomalous magnetic moment of the
nucleon \cite{BFHM}.
The situation is somewhat different for the radii: While we find
that approximately 22\% of the squared radius of $G_2$ originate from
short range physics \cite{GH}, the translation into
the Jones-Scadron basis drastically amplifies this contribution to the
radii of the quadrupole form factors
\footnote{From
eq.(\ref{defr}) one can see, that the radii
are normalized to the the size of the respective form factors at $Q^2=0$. The
above statements result from just separating the radii into long-
and short range physics and keeping the full values for $G_i(Q^2=0)$ given in table
\ref{tab:radii}.}:\\
\begin{eqnarray}
r_{M,\textnormal{Re}}^2 & =
&\left(\left.0.650\right|_{\textnormal{pc}}-\left.0.023\right|_{\textnormal{sd}}\right)\textnormal{fm}^2,
\label{sdr1}\\
r_{E,\textnormal{Re}}^2 & = &\left(\left.1.31\right|_{\textnormal{pc}}-\left.2.15\right|_{\textnormal{sd}}\right)\textnormal{fm}^2,\\
r_{C,\textnormal{Re}}^2 & =
&\left(\left.-0.019\right|_{\textnormal{pc}}-\left.0.710\right|_{\textnormal{sd}}\right)\textnormal{fm}^2.
\label{sdr2}
\end{eqnarray}
All parts marked as short distance contributions in eqs.
(\ref{sdr1})-(\ref{sdr2}) exclusively arise from the local operator
contributing to $r_{2,\textnormal{Re}}^2$. From this observation one
can see that the $G_i$ basis is clearly preferred for the
discussion of chiral signatures in the $N\Delta$-transition as there
are fewer kinematical cancellations between large numbers in this basis.
At the order of our calculation all effects beyond the linear $Q^2$-dependence of the form factors and
hence the rich structures seen in figures \ref{GM_err}-\ref{CMR_err}
exclusively originate from pion cloud dynamics.\\
In addition to these results ChEFT provides us with the knowledge
about the structures which can arise in a calculation of the form factors at
higher orders. The additional structures contributing
to each form factor at lowest order beyond our calculation read:
\begin{eqnarray}
G_1^{(3)}(Q^2) & \rightarrow & G_1^{(3)}(Q^2)+\frac{Q^2}{(4\pi
F_{\pi})^2}\delta_1,\\
G_2^{(3)}(Q^2) & \rightarrow & G_1^{(3)}(Q^2)+\frac{Q^4}{(4\pi
F_{\pi})^2 M_N^2}\delta_2,\\
G_3^{(3)}(Q^2) & \rightarrow & G_3^{(3)}(Q^2)+\frac{M_N \Delta}{(4\pi
F_{\pi})^2}\delta_3.
\end{eqnarray}
Here only those contributions which can not be absorbed via a
reparametrization of the three free parameters of our calculation
where considered. The uncertainties
when neglecting the above given structures in the
$\mathcal{O}(\epsilon^3)$ calculation are estimated by varying the
coefficients of these structures within their natural size,
i.e. between $-3$ and $3$. A stronger constraint is put on the value of $\delta_1$
which dominates the error of the magnetic dipole form factor,
since we demand the result to be consistent with the input data
of our analysis (i.e. $\mathcal{G}_M^*(Q^2)$ at low $Q^2$). This
condition is only fulfilled for $0<\delta_1<2$.\\
The solid lines in figures \ref{GM_err}, \ref{EMR_err} and
\ref{CMR_err} represent the outcome of our order $\epsilon^3$ SSE
analysis for $\mathcal{G}_M^*(Q^2)$, EMR$(Q^2)$ and CMR$(Q^2)$,
respectively. The gray shaded band around these curves marks the area
in which the array of curves with parameters $0<\delta_1<2$, $-3<\delta_2<3$ and
$-3<\delta_3<3$ lies. This band indicates the uncertainties which arise
when neglecting
higher order effects. A further source of errors lies in the
values of those low energy constants already present in the
$\mathcal{O}(\epsilon^3)$ calculation, which have been kept fixed in
the error analysis presented in figures \ref{GM_err}, \ref{EMR_err} and
\ref{CMR_err}. This is the reason for the fact that the shown error
bands for $\mathcal{G}_M^*(Q^2)$ and EMR$(Q^2)$ shrink to zero for
$Q^2\rightarrow 0$. As the quality of the determination of the low
energy constants depends on
the quality of the experimental data used as input, the typical experimental
errors of each quantity indicate the possible variation of the ChPT result.
The conclusion from the here presented error analysis is, that the calculation at
leading one loop order gives a trustworthy prediction of all three transition
form factors for momentum transfer $Q^2$ smaller than $0.2$GeV$^2$
(note that neither the $Q^2$ dependence of EMR nor any information
about CMR was used as input for our extraction of the low energy
constants; the given curves for the quadrupole moments are a prediction). Beyond
$Q^2=0.2$GeV$^2$ higher order effects can -- according to this
analysis -- play a
decisive role.
We want to emphasize that due to the uncertainties arising from the extraction of the low
energy constants the shown results for the quadrupoles are not in
contradiction with most of the models shown in the same figures.
E.g., if we where to use the  $Q^2=0$-values for the quadrupole moments
from Sato
and Lee \cite{SL} as input for our analysis (instead of the
experimental EMR$(0)$), the SSE result would exactly agree with this model prediction
(see figure \ref{CMR_err}).
Furthermore, we observe that all models (Sato-Lee \cite{SL} and DMT
\cite{DMTMAID}) and calculations (our analysis, its predecessor
\cite{GHKP} and the calculation in the $\delta$-scheme \cite{PV})
containing pion cloud effects coincidingly predict a decreasing EMR at
low $Q^2$ (where pion cloud effects should be relevant).

\section{Acknowledgements}
We thank Aron M.  Bernstein ans Costas N. Papanicolas, the organizers of this
workshop, for giving us the opportunity to present our results. This
research is part of the EU Integrated Infrastructure Initiative Hadron
Physics under contract number RII3-CT-2004-506078.


\begin{thebibliography}{99}

\bibitem{GH}
T.A. Gail and T.R. Hemmert, Eur. Phys. J. A28:91-105 (2006).

\bibitem{GHKP} G.C. Gellas, T.R. Hemmert, C.N. Ktorides and G.I. Poulis,
	Phys. Rev. D60:054022 (1999).

\bibitem{JS} H.F. Jones and M.D. Scadron, Annals Phys.81:1-14 (1973).


\bibitem{OOPS} N.F. Sparveris et al. (OOPS Collaboration),
Phys. Rev. Lett. 94:022003 (2005).

\bibitem{PDG} S. Eidelman et al. (Particle Data Group),
Phys. Lett. B592:1 (2004). 


\bibitem{BKKM} V. Bernard, N. Kaiser, J. Kambor, U.-G. Mei{\ss}ner,
Nucl. Phys. B388:315-345 (1992).


\bibitem{HHK} T.R. Hemmert, B.R. Holstein and J. Kambor,. J. Phys. G24:1831-1859 (1998).


\bibitem{PV} V. Pascalutsa and M. Vanderhaeghen,
Phys. Rev. D73:034003 (2006). 

\bibitem{d-scheme}
V. Pascalutsa and D.R. Phillips, Phys. Rev. C67:055202 (2003). 


\bibitem{BFHM}
 V. Bernard, H.W. Fearing, T.R. Hemmert and U.-G. Mei{\ss}ner,
Nucl. Phys. A635:121-145 (1998), Erratum-ibid. A642:563-563 (1998) and
T.R. Hemmert and W. Weise, Eur. Phys. J. A15:487-504 (2002). 


\bibitem{KM}
J. Gasser, M.E. Sainio, A. Svarc, Nucl. Phys. B307:779 (1988) and  
B. Kubis and U.-G. Mei{\ss}ner, Nucl. Phys. A679:698-734 (2001).

\bibitem{Alexandrou}
C. Alexandrou, et al., Phys. Rev. Lett. 94:021601 (2005).


\bibitem{SL}
 T. Sato and T.S.H. Lee, Phys. Rev. C63:055201 (2001) and 
 T. Sato and T.S.H. Lee, Phys. Rev. C54:2660-2684 (1996).


\bibitem{DMTMAID} S.S. Kamalov, et al., Phys. Rev. Lett. 83:4494 (1999) and Phys. Rev. C 64:032201 (2001).\\
D. Drechsel, O. Hanstein, S.S. Kamalov and L. Tiator,
Nucl. Phys. A645:145 (1999).\\
\texttt{http://www.kph.uni-mainz.de/MAID}

\bibitem{Beck} R. Beck et al., Phys. Rev. C61:035204 (2000). 

\bibitem{Stein} S. Stein et al., Phys. Rev. D12:1884 (1975).

\bibitem{rean} L. Tiator, D. Drechsel, S.S. Kamalov and S.N. Yang,
Eur. Phys. J. A17:357-363 (2003).

\bibitem{Baetzner} K. B\"atzner et al., Phys. Lett. B39:575-578
(1972).

\bibitem{Bartel} W. Bartel et al., Phys. Lett. B28:148-151 (1968).

\bibitem{Stave} S. Stave et al. {\tt [nucl-ex/0604013]}.

\bibitem{MAMI} Th. Pospischil et al., Phys. Rev. Lett. 86:2959-2962
(2001) and D. Elsner et al, Eur. Phys. J. A27:91-97 (2006).

\bibitem{Bonn} R. Siddle et al., Nucl. Phys. B35:93-119 (1971) and
K. Baetzner et al., Nucl. Phys. B76:1-14 (1974).

\end{thebibliography}
\end{document}